\documentstyle[aps,prb,psfig]{revtex}
\draft
\begin{document}
\twocolumn[\hsize\textwidth\columnwidth\hsize\csname
@twocolumnfalse\endcsname
\title{Can pulling cause right- to left-handed structural  transitions in 
 negatively supercoiled DNA double-helix?}
\author{Zhou Haijun{\cite{zhouhj}}}
\address{Institute of Theoretical Physics,
 Academia Sinica, 
P.O. Box 2735, Beijing 100080, China \\
State Key Lab. of Scientific and Engineering Computing, 
Beijing 100080, China}
\author{Ou-Yang Zhong-can}
\address{Institute of Theoretical Physics, Academia Sinica,
P.O. Box 2735, Beijing 100080, China}

\date{\today}
\maketitle

\begin{abstract}
The folding angle distribution of stretched and negatively supercoiled
DNA double-helix is investigated based on a previously proposed  
model of double-stranded
biopolymers (H. Zhou  {\it et al.}, Phys. Rev. Lett. {\bf 82}, 4560
 (1999)).  
It is shown that pulling can transit a negatively supercoiled DNA  double-helix from the
right-handed  B-form to a left-handed configuration which resembles
DNA  Z-form in some important respects.  
The energetics of this possible transition is calculated and the comparison
with recent experimental observations is qualitatively discussed.
\end{abstract}

\vskip2pc]
\bigskip

\newpage

Because of its vital biological significance, the elasticity of
DNA has invoked considerable  interest during  recent years,
and it is now known experimentally that  radical transitions in
DNA internal structure can be induced by the action of mechanical
 forces and/or torques. For example, pulling a DNA chain
with a force of $70$ piconewtons
(pN) will  convert   DNA standard  B-form  conformation into  a 
over-stretched  S-form \cite{cluzel96,smith96};
 and at the joint action of a positive
torque and a  force about $3$ pN,  a DNA will take on 
 a novel P-form  with exposed bases\cite{allemand98}.
Here we suggest a further possibility for this transition {\it scenario}
and show that under-twisted\cite{strick96} (negatively supercoiled) DNA
 can take on a left-handed  configuration under the action of a moderate 
stretching force. Such a  left-handed  conformation
is found to resemble  DNA Z-form\cite{watson87} in some important respects.
The energetics of this possible transition is calculated and a
qualitative comparison with very recent experiments\cite{strick98,leger99}
is also performed.

\begin{figure}
\leftline{\psfig{file=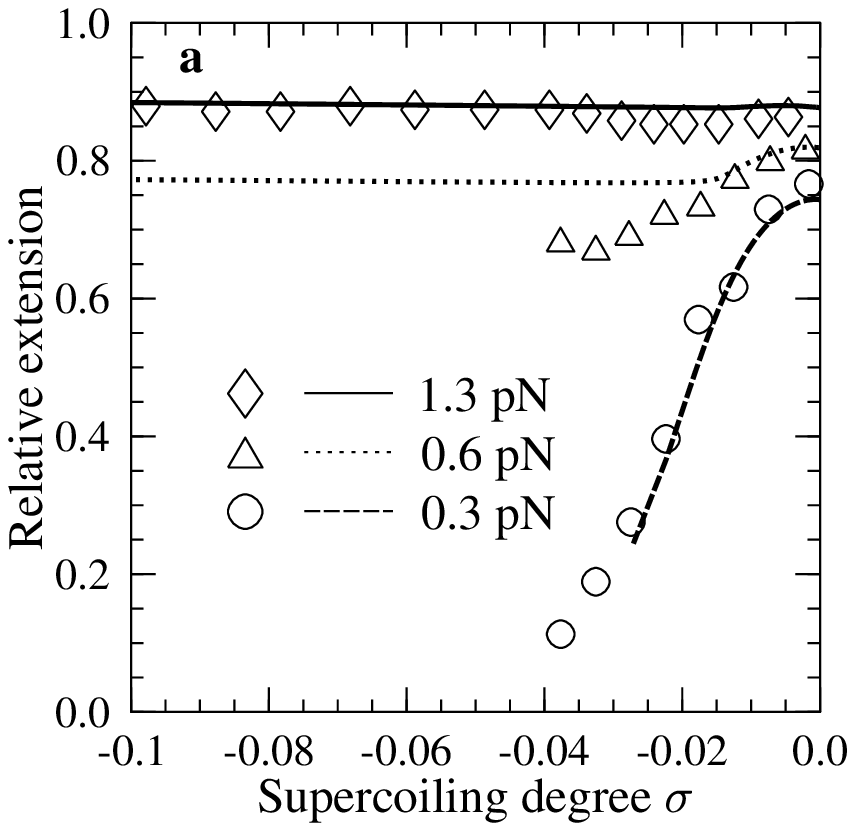,height=7.5cm}}
\vspace*{-1.0cm}
\leftline{\psfig{file=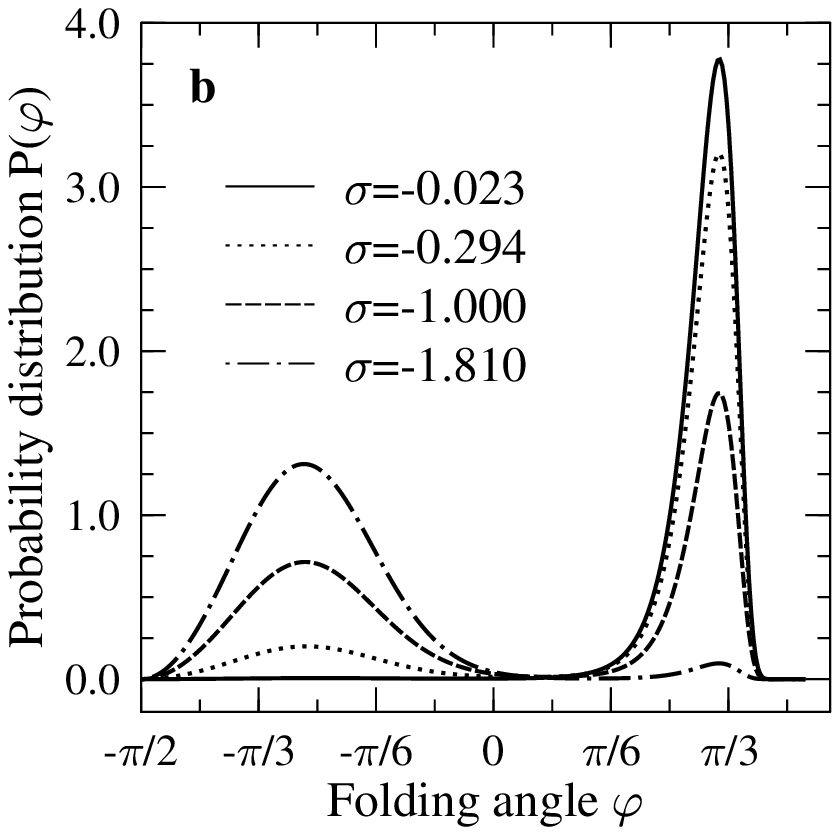,height=7.5cm}}
\caption
{{\bf a}, DNA extension vs supercoiling degree.
The supercoiling degree $\sigma$ is defined as $\sigma=
({\it lk}-{\it lk}_0)/{\it lk}_0$, where ${\it lk}_0$ and 
${\it lk}$ are, respectively, the linking number  
of  a relaxed  and a torsionally constrained DNA\cite{zhou99}.
Experimental data (symbols) are from Ref. 4.
{\bf b}, folding angle distributions for 
 a DNA pulled with a force of $1.3$ pN. Here,
the folding angle $\varphi$ is defined  in the
range from $-\pi/2$ to $\pi/2$, with negative (positive) values corresponding
to left (right)-handed configurations\cite{zhou99}. } 
\label{fig1} 
\end{figure}

The extension vs supercoiling relation\cite{strick96} for under-twisted DNA
  is studied based on a  model 
 proposed earlier\cite{zhou99} (see Fig. \ref{fig1}a).
 For small pulling forces ($\leq 0.3$ pN), a supercoiled DNA  can
 shake off its twisting stress by
 writhing its central axis and forming plectonemic
 structures\cite{marko95}. However, this  leads to shortening of DNA end-to-end
distance and hence becomes very unfavorable as the  force increases.
At this stage, the torsional stress caused by supercoiling 
begins to unwind the B-form double-helix and triggers the transition
of DNA  internal structure, where  a continuously increasing portion of 
DNA takes on some certain new configuration as supercoiling increases,
while its total extension keeps almost invariant.  

Information about the  new configuration can be revealed by
the folding angle\cite{zhou99} $\varphi$ distributions $P(\varphi)$.
 This distribution is calculated by 
\begin{equation}
P(\varphi)=\int \Phi^2({\bf t},\varphi)d {\bf t},
\label{eq1}
\end{equation}
 where ${\bf t}$ is the tangent vector of DNA's central axis and
$\Phi({\bf t}, \varphi)$ is the (normalized) ground-state eigenfunction of the
Green equation Eq. (9) in Ref. 8. 
The folding angle distribution (Fig. \ref{fig1}b) has the following aspects:
When the torsional stress is small (with the supercoiling degree
$|\sigma| < 0.025$), the distribution has only one narrow and steep peak
at $\varphi \simeq + 57.0^\circ$, indicating that  DNA is
completely in  B-form.
 With the increase of  torsional 
stress, however, another peak appears at $\varphi \simeq
- 48.6^\circ$ and the total probability for the folding angle to be 
negatively-valued increases   gradually with supercoiling.
Since negative folding angles correspond to left-handed configurations
\cite{zhou99}, we can conclude that, with the increasing of supercoiling, 
left-handed DNA conformation is nucleated and it
 then elongates along the DNA chain
as B-DNA disappears gradually. The whole chain becomes completely
left-handed at $\sigma\simeq
-1.85$.

It is worth noticing that, (i) as the supercoiling degree 
changes, the positions of the two peaks of the folding angle distribution
remain almost fixed and, (ii) between these two peaks, there exists an extended region
of folding angle from $0$ to $\pi/6$ which always has only extremely small
probability of occurrence. Thus, a negatively  supercoiled DNA 
can have two possible stable configurations, a right-handed B-form  
and a left-handed configuration with an average 
folding angle $\simeq -48.6^\circ$.
A transition between these two structures for a DNA segment
 will generally lead to an abrupt and finite variation in the folding angle.

\begin{figure}
\leftline{\psfig{file=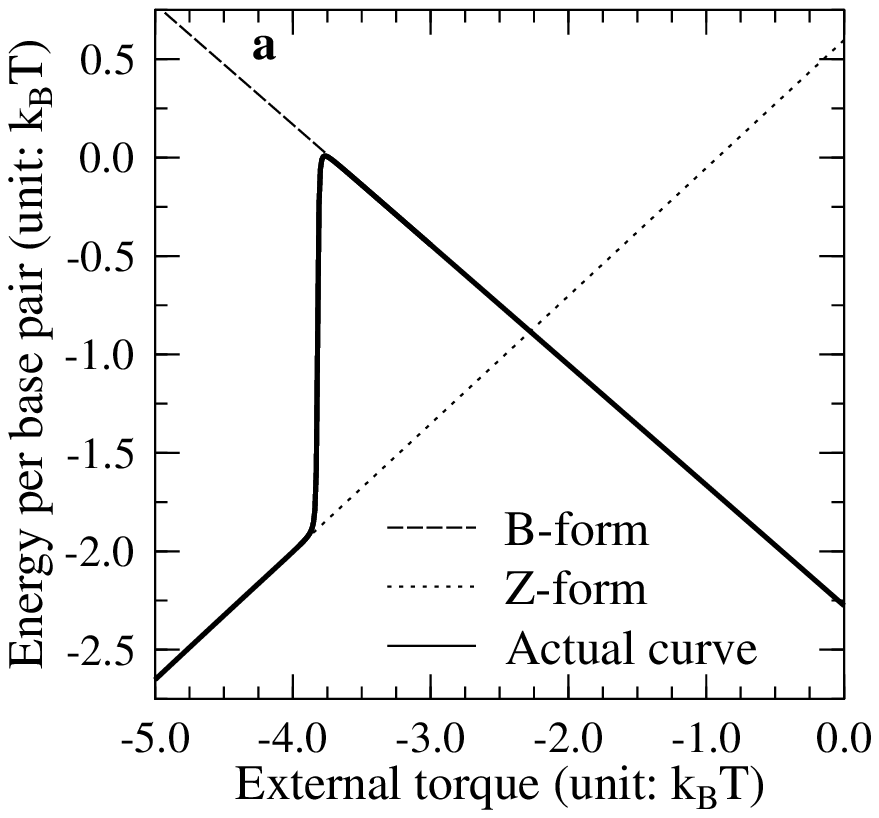,height=7.5cm}}
\vspace*{-1.0cm}
\leftline{\psfig{file=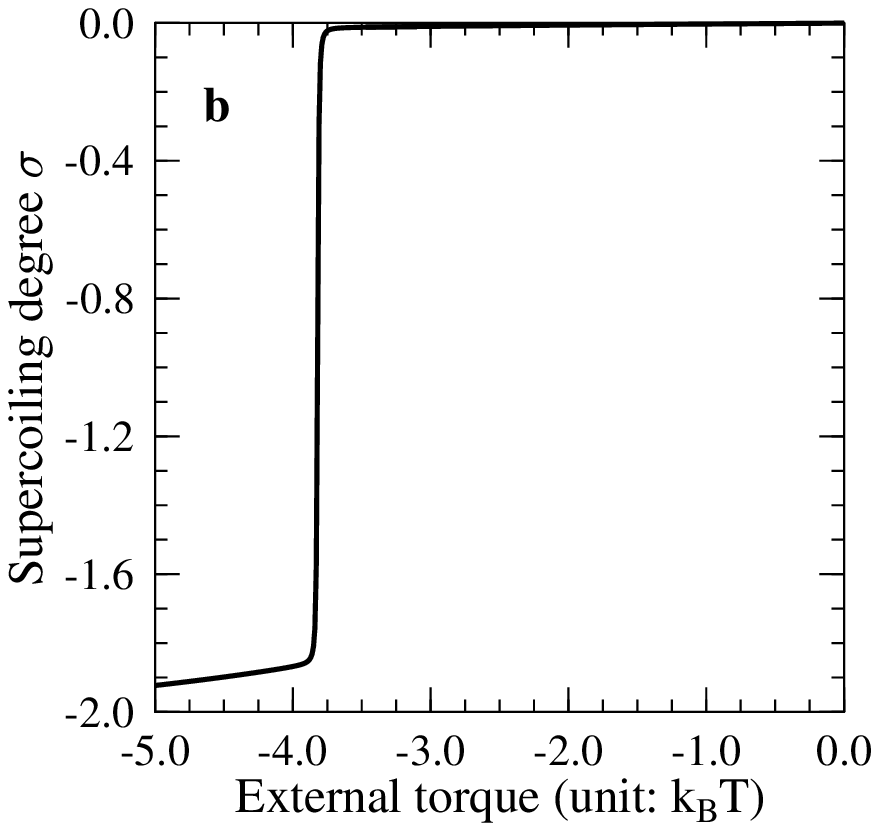,height=7.5cm}}
\caption
{{\bf a}, The sum of average base-stacking and torsional energy per
base pair  at force $1.3$ pN.
For highly extended DNA only these two  interactions
are sensitive with torque. The torque-related torsional energy density
is\cite{zhou99} $-\Gamma \sin\varphi/R$, where $R$ is DNA molecular
radius. 
{\bf b}, The relation between DNA supercoiling degree and external
torque at force  $1.3$ pN.
}
\label{fig2}
\end{figure}

The sum of the average base-stacking energy and torsional energy caused by
external torque (see caption of Fig. \ref{fig2}a)
  as   a function of torque  $\Gamma$ is shown in Fig. \ref{fig2}a
and the relation between $\sigma$ and $\Gamma$
  in Fig. \ref{fig2}b. From
these figures we can infer that, (i)
 for negative torque  less than the
critical value $\Gamma_c\simeq -3.8$ 
k$_B$T, DNA can only stay in B-form state; 
(ii) near  $\Gamma_c$  DNA can either
be right- or be left-handed and, as negative
supercoiling increases (see Fig. \ref{fig2}b) more and more DNA segments
will stay in the left-handed form, which is much lower in
energy ($\simeq -2.0$ k$_B$T per base pair (bp))
 but stable only  when torque reaches $\Gamma_c$;
(iii) for negative torque greater than $\Gamma_c$ DNA is    
completely left-handed.
B-form DNA  at $\Gamma_c$ has energy  about $0.0$ k$_B$T per bp,
indicating that the work done by external torque just
cancels the base-stacking energy. 
Therefore, it might not be enough to further break the
hydrogen bonds between DNA complementary bases and cause
denaturation \cite{strick96,strick98}.
Nevertheless, since the transition from right- to
left-handed structure requires  radical rearrangement
of DNA base pairs, the possibility of transient denaturation in
DNA double-helix can not be ruled out. This
is a subtle question, and  maybe transient denaturations can occur in
the weaker AT-rich regions, or even be {\it induced} and
then {\it captured} by the added homologous single-stranded DNA probes
 in the solution\cite{strick98}.

For the  left-handed state revealed by  Figs. \ref{fig1}b and \ref{fig2} 
we have obtained that, at $\Gamma=-4.0$ k$_B$T (where
DNA is completely  left-handed) the average rise per bp
 is about $3.83$ $\AA$, and
the pitch per turn of helix is  $46.76$ $\AA$, with the 
number of bps per turn of helix being  $12.19$.
Notice  these characteristic quantities are very similar 
with those of DNA left-handed Z-form, which are $3.8$ $\AA$,
$45.6$ $\AA$,  and $12$  bps, respectively\cite{watson87}.
We suspect that
the identified left-handed  configuration belongs to DNA Z-form.  
Recently, L\'{e}ger {\it et al.} also pointed out that Z-form structure 
should be included   to qualitatively  interpret their experimental
result\cite{leger99}.

\end{document}